\documentclass[epj,nopacs]{svjour}
\pdfoutput=1
\usepackage{graphicx,hyperref,wrapfig,cite,color,subfigure,amssymb,amsmath,epsfig}
\usepackage[utf8]{inputenc}
\usepackage[english]{babel}
\hypersetup{
     colorlinks=true,        
     linkcolor=red,          
     citecolor=green,        
     filecolor=magenta,      
     urlcolor=blue           
}
\usepackage[flushleft]{threeparttable}
\usepackage{kantlipsum}
\usepackage[usenames,dvipsnames]{xcolor}
\usepackage{soul}
\usepackage{float,caption}
\usepackage{mathtools,cuted}
\newcommand{\newc}{\newcommand}
\def\u#1{\verb!#1!\endgroup}
\newc{\HW}{\mbox{\textsf{HERWIG}}}
\newc{\Hw}{\mbox{\textsf{Herwig}}}
\newc{\KrkNLO}{\textsf{KrkNLO}}
\newc{\OL}{\textsf{OpenLoops}}
\newc{\Collier}{\textsf{Collier}}
\newc{\GoSam}{\textsf{GoSam}}
\newc{\TAUOLA}{\textsf{TAUOLA}}
\newc{\ThePEG}{\textsf{ThePEG}}
\newc{\boost}{\textsf{BOOST}}
\newc{\HepMC}{\textsf{HepMC}}
\newc{\Rivet}{\textsf{Rivet}}
\newc{\lhapdf}{\textsf{LHAPDF}}
\newc{\HWPP}{\mbox{\textsf{Herwig++}}}
\newc{\evt}{\textsf{EvtGen}}
\newc{\fortran}{\textsf{FORTRAN}}
\newc{\decayer}{\textsf{Decayer}}
\newc{\matchbox}{\textsf{Matchbox}}
\newc{\PythiaI}{\textsf{P8I}}
\newc{\Pythia}{\textsf{Pythia 8}}
\newc{\HADML}{\textsf{HADML}}

\newc{\HWPPClass}[1]{\href{https://herwig.hepforge.org/doxygen/classHerwig_1_1#1.html}{\textsf{#1}}}
\newc{\ThePEGClass}[1]{\href{https://thepeg.hepforge.org/doxygen/classThePEG_1_1#1.html}{\textsf{#1}}}
\newc{\HWPPParameter}[2]{\href{https://herwig.hepforge.org/doxygen/#1Interfaces.html\##2}{{\bf #2}}}
\newc{\ThePEGParameter}[2]{\href{https://thepeg.hepforge.org/doxygen/#1Interfaces.html\##2}{{\bf #2}}}
\newc{\HWPPParameterValue}[3]{\href{https://herwig.hepforge.org/doxygen/#1Interfaces.html\##2}{{\bf [#2=#3]}}}
\newc{\HWPPParameterValueB}[3]{\href{https://herwig.hepforge.org/doxygen/#1Interfaces.html\##2}{{\bf [#3]}}}
\newc{\ThePEGParameterValue}[3]{\href{https://thepeg.hepforge.org/doxygen/#1Interfaces.html\##2}{{\bf [#2=#3]}}}

\preprint{
CERN-TH-2023-223\\
HERWIG-2023-01\\
KA-TP-28-2023\\
MCnet-23-19\\
IPPP/23/66
}

\title{Herwig 7.3 Release Note}

\author{
 Gavin~Bewick\inst{1}\and
 Silvia~Ferrario~Ravasio\inst{2}\and
 Stefan~Gieseke\inst{3}\and
 Stefan~Kiebacher\inst{3}\and
 Mohammad~R.~Masouminia\inst{1}\and
 Andreas~Papaefstathiou\inst{4}\and
 Simon~Pl\"atzer\inst{5}\and
 Peter~Richardson\inst{1}\and
 Daniel~Samitz\inst{6}\and
 Michael~H.~Seymour\inst{7}\and
 Andrzej~Si\'odmok\inst{8}\and
 James~Whitehead\inst{9}
}
\institute{
  IPPP, Department of Physics, Durham University,\and
  CERN, Theoretical Physics Department, Geneva,\and
  Institute for Theoretical Physics, Karlsruhe Institute of Technology,\and
 Department of Physics, Kennesaw State University, 830 Polytechnic Lane, Marietta, GA 30060, USA\and
  Theoretical Physics, NAWI Graz, University of Graz and
  Particle Physics, Faculty of Physics, University of Vienna,\and
  Stefan Meyer Institute for Subatomic Physics, Austrian Academy of Sciences, 1010 Wien, Austria, \and
  Particle Physics Group, Department of Physics and Astronomy, University of Manchester,\and
  Jagiellonian University, ul. prof. Stanisława Łojasiewicza 11, 30-348 Kraków, 
  Poland, \and
  The Henryk Niewodniczański Institute of Nuclear Physics in Cracow, Polish
  Academy of Sciences.  
}

\date{December 8, 2023}

\abstract{A new release of the Monte Carlo event generator
  \textsf{Herwig} (version 7.3) has been launched. This iteration
  encompasses several enhancements over its predecessor, version
  7.2. Noteworthy upgrades include: the implementation of a
  process-independent electroweak angular-ordered parton shower
  integrated with QCD and QED radiation; a new recoil scheme for
  initial-state radiation improving the behaviour of the
  angular-ordered parton shower; the incorporation of the heavy quark
  effective theory to refine the hadronization and decay of excited
  heavy mesons and heavy baryons; a dynamic strategy to regulate the
  kinematic threshold of cluster splittings within the cluster
  hadronization model; several improvements to the structure of the
  cluster hadronization model allowing for refined models; the
  possibility to extract event-by-event hadronization corrections in a
  well-defined way; the possibility of using the string model, with a
  dedicated tune. Additionally, a new tuning of the parton shower and
  hadronization parameters has been executed. This article discusses
  the novel features introduced in version 7.3.0.  }

\begin{document}\sloppy

\maketitle

\section{Introduction}

\Hw, a multi-purpose event generator for reactions at $pp$, $ep$ and
$ee$ colliders, is now available in a new version, $7.3$. The \Hw~7
release series, started with \cite{Bellm:2015jjp} and continuing with
\cite{Bellm:2017bvx,Bellm:2019zci}, is based on the \HWPP\ development
\cite{Bahr:2008pv,Bahr:2008tx,Bahr:2008tf,Gieseke:2011na,Arnold:2012fq,Bellm:2013lba}
and fully supersedes the previous \HWPP\ and \HW\ versions.  The main
cornerstones of the \Hw~7 series are significant improvements to the
prediction of the hard scattering, now routinely available at
next-to-leading order QCD, as well as major development and
theoretical insight into the available angular-ordered and dipole
parton shower algorithms and their accuracy. A major follow-up
release of these previous \Hw~7 versions is now available as
\Hw~7.3. The updates in this release are centred around parton shower
algorithms, the hadronization model and soft QCD interactions in
hadronic collisions. Please refer to the \HWPP\ manual
\cite{Bahr:2008pv}, the \Hw~7.0 \cite{Bellm:2015jjp} as well as this
release note when using the new version of the program.  Studies or
analyses that rely on a particular feature of the program should also
reference the paper(s) where the physics of that feature was first
described.  The authors are happy to provide guidance on which
features are relevant for a particular analysis. A major publication
intended as physics reference and manual will be appearing in the
coming year.

\subsection{Availability}

The new version, as well as older versions of the \Hw\ event generator can be
downloaded from the website
\texttt{\href{https://herwig.hepforge.org/}{https://herwig.hepforge.org/}}.
We strongly recommend using the \texttt{bootstrap} script provided for the
convenient installation of \Hw\ and all of its dependencies, which can be
obtained from the same location. On the website, tutorials and FAQ sections are
provided to help with the usage of the program. Further enquiries should be
directed to \texttt{herwig@projects.hepforge.org}.  \Hw\ is released under the
GNU General Public License (GPL) version 3 and the MCnet guidelines for the
distribution and usage of event generator software in an academic setting, see
the source code archive or
\texttt{\href{http://www.montecarlonet.org/index.php?p=Publications/Guidelines}
  {http://www.montecarlonet.org/}}.

\subsection{Prerequisites and Further Details}

\Hw~7.3 is built on the same backbone and dependencies as its
predecessors \Hw~7.0, 7.1 and 7.2, and uses the same method of build,
installation and run environment. No major changes should hence be
required in comparison to a working \Hw~7.2 installation. Some of the
changes, though, might require different compiler versions.  The
tutorials at
\texttt{\href{https://herwig.hepforge.org/tutorials/}{https://herwig.hepforge.org/tutorials/}}
have been extended and adapted to the new version and serve as the
primary reference for physics setups and as a user manual until a
comprehensive replacement for the detailed manual \cite{Bahr:2008pv}
is available.

In \Hw~7.3, one backward incompatibility has been introduced that will
affect users' input files. The previous instance of QCD coupling,
\texttt{AlphaQCD}, has been replaced by two instances.
\texttt{AlphaQCDFSR} is used for processes with only final-state
radiation, i.e.~final-state parton showers, matrix element corrections
to $e^+e^-$ processes and partonic decays of final-state objects.
\texttt{AlphaQCDISR} is used for all other processes, including
initial-state parton showers and built-in matrix elements for hadron
collisions. Users should change instances of \texttt{AlphaQCD} in their
input files to \texttt{AlphaQCDISR}, unless they are specifically
targeted at $e^+e^-$ or other final-state effects such as top quark
decays, in which case they should use \texttt{AlphaQCDFSR}.

\section{Improvements on the angular ordered parton shower}

Process-independent parton showers have long been one of the pivotal
components in all multi-purpose event generators for particle
physics. One of the key features of \Hw~7 has been the provision of two
complementary paradigms for parton showering: the angular ordered parton
shower and the transverse momentum ordered dipole shower. Version~7.3
includes significant improvements to the angular ordered parton shower,
including electroweak radiation for the first time, and improving the
method by which partons' kinematics are reconstructed at the end of the
shower (the recoil scheme).

Predominantly, the current paradigm for parton showers revolves around the QCD+QED schemes, which, while yielding satisfactory results in present conditions, might not be sufficient as we approach higher energy scales where it is anticipated that EW bosons will begin to manifest as massless partons, a prediction corroborated by recent LHC observations~\cite{ATLAS:2015egz, ATLAS:2018jvf}. Further compelling evidence for this shift is found in the corresponding electroweak virtual corrections, which are both substantial and predominantly negative. These observations and findings underline the pressing need to innovate beyond the current status quo. There exists a strong justification for the introduction of a process-independent EW parton shower, essentially pushing the envelope and evolving the parton shower framework to a more comprehensive QCD+QED+EW scheme~\cite{Masouminia:2021kne}.

The \Hw{} angular-ordered parton shower dresses the hard event with a
series of iterated $1\to 2$ ordered branchings.
When a new emission is generated, it is necessary to perform momentum
reshuffling to assign virtuality to the pre-branching emitter,
allowing the splitting.
This procedure makes it impossible to preserve all the kinematic
invariants associated with previous branchings.
The choice of the preserved variable defines the recoil scheme.
Since the ordering scale can be expressed in terms of the preserved
variable, the recoil scheme coincides with the interpretation of
the ordering variable.
\Hw~7.2 featured three recoil schemes for final-state radiation~(FSR):
the transverse-momentum preserving scheme~\cite{Gieseke:2003rz}, the
virtuality-preserving scheme~\cite{Reichelt:2017hts}, and the
so-called ``dot-product'' preserving scheme~\cite{Bewick:2019rbu}.
This latter scheme preserves the dot product between the momenta of the
post-branching partons originating from the splitting.
Initial-state radiation~(ISR) was always reconstructed using the
transverse-momentum recoil scheme~\cite{Gieseke:2003rz}.
Version~7.3 enables the consistent use of all three schemes for FSR
and ISR~\cite{Bewick:2021nhc}.
The dot-product preserving scheme is now the default for both ISR and FSR.

\subsection{Process-independent electroweak radiation}

Using the fundamental shower kinematics and dynamics of \textsf{Herwig~7} in the \textit{quasi}-collinear limit~\cite{Catani:2000ef,Bahr:2008pv,Masouminia:2021kne} while assuming a generic splitting $\widetilde{ij} \to i + j$ for the quark splittings:
\begin{align}
q\to q'W^{\pm}, \quad q\to qZ^{0},
\label{qqV}
\end{align}
\begin{align}
q\to qH,
\label{qqH}
\end{align}
as well as the EW gauge boson splittings:
\begin{align}
W^{\pm} \to W^{\pm}Z^{0}, \quad & W^{\pm} \to W^{\pm} \gamma, \quad
Z^{0} \to W^{+}W^{-}, 
\label{VVV}
\nonumber \\
&\gamma \to W^{+}W^{-}, 
\end{align}
\begin{align}
W^{\pm} \to W^{\pm} H, \quad Z^{0} \to Z^{0} H,
\label{VVH}
\end{align}
one can write the helicity amplitudes of $\widetilde{ij} \to i + j$ in the \textit{quasi}-collinear limit as:
\begin{equation}
H_{\widetilde{ij} \to i + j}(z,\tilde{q};\lambda_{\widetilde{ij}},\lambda_i,\lambda_j) = g\sqrt{\frac2{\tilde{q}^2_{\widetilde{ij}}-m_{\widetilde{ij}}^2}} F_{\lambda_{\widetilde{ij}},\lambda_i,\lambda_j}^{\widetilde{ij} \to i + j},
\end{equation}
where $\widetilde{q}$ denotes the evolution scale of the shower, and $m_{\widetilde{ij}}$ the mass of progenitor. Additionally, the vertex function $F_{\lambda_{\widetilde{ij}},\lambda_i,\lambda_j}^{\widetilde{ij} \to i + j}$ is derived exclusively from Feynman rules~\cite{Masouminia:2021kne}. Consequently, the splitting function can be delineated as:
\begin{equation}
P_{\widetilde{ij} \to i + j}(z,\tilde{q}) = \sum_{\mathrm{spins}} \left| H_{\widetilde{ij} \to i + j}(z,\tilde{q}; \lambda_{\widetilde{ij}},\lambda_i,\lambda_j) \right|^2,
\label{QSF}
\end{equation}
and explicit expressions for Eq.~\eqref{QSF} can be given for the splittings \eqref{qqV}$-$\eqref{VVH} as:
\begin{align}
	&P_{q\to q' V}(z,\tilde{q}) =
	\frac {1}{1-z} \bigg[ \big( {g_L}^2\rho_{{-1,-1}}+{g_R}^2\rho_{{1,1}} \big)
	\nonumber \\
	&\;\; \Big\{ \big(1+z^2 \big) \big(1+\frac {m_0^2}{\tilde{q}^2z\big(1-z\big)}\big)
	-\frac{m_1^2\big(1+z\big)}{z\tilde{q}^2\big(1-z\big)}
	\Big.  
	\nonumber\\ 
	&\;\; - \Big.
	\frac {m_2^2}{z\tilde{q}^2} \Big\}  
	+ \frac {{m_0}^2}{\tilde{q}^2} \big( {g_L}^2\rho_{{1,1}}+{g_R}^2\rho_{{-1,-1}}\big)
	\nonumber \\
	&\;\; -\frac {2m_0m_1}{z\tilde{q}^2} g_Lg_R\big(\rho_{{1,1}}+\rho_{{-1,-1}} \big) \bigg],
\end{align}
\begin{align}
	P_{q\to qH}(z,\tilde{q}) &= g^2 ({m_0 \over m_W})^2 \left[ (1-z)+
	\frac{4 m_0^2 - m_2^2 }{\tilde{q}^2 (1-z) z}
	\right],
\end{align}

\begin{align}
	&P_{V \to V'V''}(z,\tilde{q}) = 
	\frac{2 \big( \rho_{-1,-1}+\rho_{1,1} \big)}{(1-z) z} \Bigg[ \big(1-(1-z) z\big)^2 
	\nonumber \\
	&\;\; + m_{0,t}^2 \big(1-(1-z) z\big)^2 - m_{1,t}^2 \big(1-(1-z) z^2\big) 
	\nonumber \\
	&\;\; - m_{2,t}^2 \big(1-(1-z)^2 z\big) + 2 \rho_{0,0} m_{0,t}^2 z (1-z)^3 \Bigg],
\end{align}
\begin{align}
&P_{V\to VH}(z,\tilde{q}) = 
\frac{1-z}{4 z} \Big[ z^2 \big( \rho_{-1,-1} + \rho_{1,1} \big) + 2 \rho_{0,0} \Big] 
\nonumber \\
&\;\; - \frac{m_{H,t}^2 }{4 z} \Big[ z^2 \big( \rho_{-1,-1} + \rho_{1,1} \big) + 2 \rho_{0,0} \Big] \nonumber \\
&\;\; - \frac{m_{0,t}^2 }{4 z^2} \Bigg[ \Big( 2 z^2-4 z+2\Big) \rho_{0,0}
\nonumber \\
&\;\; +\Big(z^4-2 z^3-z^2 \Big) \big( \rho_{-1,-1}+\rho_{1,1} \big) \Bigg],
\end{align}
with $m_k, \; k=0,\;1,\;2$ being the masses of the particles $\widetilde{ij}, \;i, \;j$ respectively and $m_{k,t}^2 = m_k^2/(\tilde{q}^2z(1-z))$. Note that the expressions above are derived using Dawson's approach~\cite{Dawson:1984gx}. In this approach, terms in the longitudinal polarization vectors of the EW vector bosons, which are proportional to their momenta, are systematically removed to prevent irreducible divergences in the resulting splitting functions~\cite{Masouminia:2021kne}. 

On the development side, the helicity-dependent splitting kernels for \eqref{qqV}$-$\eqref{VVH} are introduced into \Hw~7's AO parton shower though the introduction of four new classes, namely \textcolor{blue}{\texttt{HalfHalfOneEWSplitFn}}, \textcolor{blue}{\texttt{HalfHalfZeroEWSplitFn}}, \textcolor{blue}{\texttt{OneOneOneEWSplitFn}} and \textcolor{blue}{\texttt{OneOneZeroEWSplitFn}} respectively. The first two can by default contribute to both initial and final state radiation, while the latter two are explicitly final state splittings. This is because the inclusion of backward EW vector boson evolution in the initial state parton shower requires the use of EW parton distribution functions, which will reduce the generality and efficiency of the algorithm and does not increase its reliability.

Additionally, each weakly interacting progenitor in the AO shower is now tagged with an EW scale, alongside QCD and QED scales, allowing for a seamless  interleaving of these interactions via the new \texttt{QCD+QED+EW} default scheme. This choice can be altered via the following interface commands
\\[0.05in]
\texttt{cd /Herwig/Shower}\\
\texttt{set ShowerHandler:Interactions <scheme>}
\\[0.05in]
Here, the option \texttt{ALL} corresponds to the \texttt{QCD+QED+EW} shower scheme. The other available options are \texttt{QEDQCD}, \texttt{QCD}, \texttt{QED} and \texttt{EWOnly}. 

The performance and physics of this new iteration of the AO parton shower has been comprehensively validated in~\cite{Masouminia:2021kne} and its phenomenology in producing credible predictions against experimental observations has been surveyed in~\cite{Darvishi:2021het,Darvishi:2020paz}. 

\begin{figure}
\centering
\includegraphics[width=.49\textwidth]{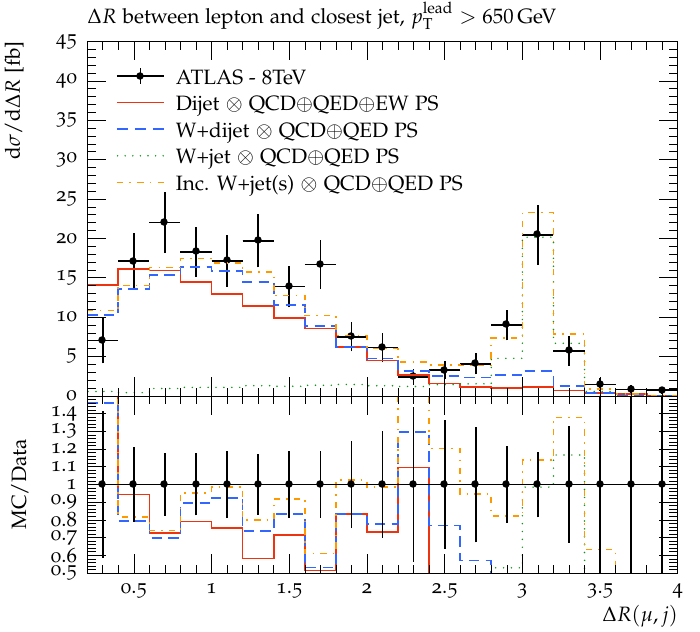}
\caption{Angular distribution of a $W$ boson in association with a leading jet possessing transverse momentum greater than 650 GeV, observed at a center-of-mass energy of 8 TeV, as detailed in~\cite{Masouminia:2021kne}.}
\label{EWPS-ATLAS-650}
\end{figure}

As an example, Figure~\ref{EWPS-ATLAS-650}~\cite{Masouminia:2021kne} shows the angular distribution of $W^{\pm}$ bosons and high-$p_t$ jets at a center-of-mass energy of 8 TeV, using ATLAS data \cite{Aaboud:2016ylh}. It specifically highlights the distribution of muons relative to the nearest jet with a transverse momentum greater than 650 GeV. The blue dashed and green dotted lines represent simulations with $W^{\pm}+jet$ and $W^{\pm}+2jets$ matrix elements (MEs), respectively, and the orange dash-dotted line is the combined result of these two MEs. These MEs were created with \textsf{MadGraph}~\cite{Alwall:2011uj} and then processed with \textsf{Herwig}'s \texttt{QCD+QED} showering scheme. The red solid line, however, is derived from a pure QCD dijet event showered using \textsf{Herwig}'s comprehensive \texttt{QCD+QED+EW} scheme. The proximity of the red histogram to the blue in this Figure showcases the effectiveness of \textsf{Herwig}'s \texttt{QCD+QED+EW} parton shower scheme in accurately reproducing the angular distribution of $W^{\pm}$ bosons with high-$p_t$ jets, even without explicit $W$ emissions from the hard process.

\subsection{Dot-product preserving recoil scheme}
The \Hw\ angular-ordered parton shower algorithm\cite{Gieseke:2003rz} is
built on the Lorentz-invariant variables $z$ and $\tilde{q}^2$, where
$z$ specifies the momentum sharing between the children in a $q_0\to
q_1+q_2$ branching:
\begin{equation}
  z=\frac{q_1\!\cdot\! n}{q_0\!\cdot\! n}\,,
  \label{eq:zdef}
\end{equation}
where $n$ is an auxiliary light-like vector specifying the collinear
direction, and $\tilde{q}$ generalizes the emitter energy times opening
angle\footnote{To simplify the presentation here, we neglect parton
masses, but they are fully implemented in the algorithm
\cite{Bewick:2019rbu} and code. We also describe the algorithm only for
final-state evolution, but again it is fully implemented also for
initial-state evolution\cite{Bewick:2021nhc}.},
\begin{equation}
  \tilde{q}^2=\frac{q_0^2}{z(1-z)}\,.
  \label{eq:q2tildedef}
\end{equation}
The algorithm works by generating a $(z,\tilde{q}^2)$ pair for a
branching, then allowing each of its children to probabilistically evolve
and generate further branching, until a termination condition is
reached. For a branching in which the children do not branch further, so
are on their mass shell, Eqs.~(\ref{eq:zdef},\ref{eq:q2tildedef}) are
unambiguous. The transverse momentum of the branching can be calculated,
\begin{equation}
  p_T^2=z^2(1-z)^2\tilde{q}^2,
  \label{eq:qTscheme}
\end{equation}
and the kinematics reconstructed.

However, Eq.~(\ref{eq:q2tildedef}) is not the unique definition
of~$\tilde{q}^2$. When the children are on-shell, it can be written in any
of the equivalent forms
\begin{equation}
  \tilde{q}^2=\frac{p_T^2}{z^2(1-z)^2}=\frac{q_0^2}{z(1-z)}=\frac{2q_1\!\cdot\!
  q_2}{z(1-z)}\,.
  \label{eq:q2tildedef2}
\end{equation}
When the children have acquired a non-vanishing virtuality through their subsequent
evolution, we can continue to use Eq.~(\ref{eq:zdef}), but we can choose
to use only one of the three definitions of $\tilde{q}^2$ in
Eq.~(\ref{eq:q2tildedef2}) and the value of $p_T^2$ that is
reconstructed for a given $\tilde{q}^2$ value will be different in each
case.

The original choice of Ref.~\cite{Gieseke:2003rz} was to use
Eq.~(\ref{eq:qTscheme}), which we now call the {\bf\boldmath
  $p_T$-preserving scheme}.

In Ref.~\cite{Reichelt:2017hts}, it was pointed out that the
$p_T$-preserving scheme gives too much hard radiation, as the
virtuality of the parent parton can grow arbitrarily after multiple
emissions. It was suggested that an alternative scheme, based on the
second of the options in Eq.~(\ref{eq:q2tildedef2}) does not suffer
this problem. We call this the {\bf\boldmath $q^2$-preserving
  scheme}. It results in
\begin{equation}
  p_T^2=z^2(1-z)^2\tilde{q}^2-q_1^2(1-z)-q_2^2z,
  \label{eq:q2scheme}
\end{equation}
which is clearly smaller when there is subsequent emission, resulting
in a less strong growth of virtuality after multiple emissions and a
better description of LEP data at
$\sqrt{s}=M_{Z^0}=91.1876$~GeV. However, it was found that this
definition sometimes results in events that should have a negative
$p_T^2$ according to this formula, so cannot be reconstructed. This
was remedied by replacing $p_T$ by zero in those events.
However, in Ref.~\cite{Bewick:2019rbu}, it was shown that the adoption
of the $q^2$-preserving scheme results in a formal loss of
logarithmic accuracy.
A third scheme, the {\bf dot-product $(q_1\cdot q_2)$ preserving
  scheme} was then defined, based on the third option in
Eq.~(\ref{eq:q2tildedef2}). It results in
\begin{equation}
  p_T^2=z^2(1-z)^2\tilde{q}^2-q_1^2(1-z)^2-q_2^2z^2.
  \label{eq:dotscheme}
\end{equation}
We can see that this is intermediate between the other two schemes and
is found to have a `best of both' behaviour: the virtuality does not
grow excessively in multiple emissions, but remains within the
kinematically reconstructible phase space and maintains the same
formal logarithmic accuracy of the $p_T$-preserving formulation, which
reaches Next-to-Leading Logarithms for global observables, such as
event shapes.

The dot-product preserving scheme was extended to initial-state
evolution in Ref.~\cite{Bewick:2021nhc}.
When an incoming parton $q_1$ backward-evolves into a parton $q_0$
(emitting a final-state parton $q_2$), the ordering variable can be
written as
\begin{equation}
\tilde{q}^2 = \frac{p_T^2}{(1-z)^2} = \frac{-q_1^2}{1-z} = \frac{2 q_0 \cdot q_2}{1-z}.
\end{equation}
It was noted that in the case of ISR, subsequent emissions tend to
increase the value of $p_T^2$ because incoming partons develop a
negative virtuality, in contrast to the FSR case, where new emissions
tend to decrease $p_T^2$.
Tuning the strong coupling constant $\alpha_s$ using the
transverse-momentum distribution of the $Z$ boson measured at the LHC,
which is primarily sensitive to ISR, results in a value larger than
the one obtained from LEP data, that instead exclusively probes FSR.
For this reason we introduced separate values of $\alpha_s$ for ISR
and FSR. 
The previous instance of \texttt{AlphaQCD} has now been replaced by two
instances, \texttt{AlphaQCDFSR} and \texttt{AlphaQCDISR}.
\texttt{AlphaQCDISR} is used for the
initial-state shower and for built-in matrix elements for hadron
collisions, while \texttt{AlphaQCDFSR} is used for the final-state
shower, matrix element corrections to built-in matrix elements for
$e^+e^-$ and for final-state decays.  Their input values (defined in the
CMW scheme\cite{Catani:1990rr} at $M_{Z^0}$) can be set by
\\[0.05in]
\texttt{cd /Herwig/Shower}\\
\texttt{set AlphaQCDISR:AlphaIn <value>}\\
\texttt{set AlphaQCDFSR:AlphaIn <value>}.

\begin{figure}[t]
\centering
\includegraphics[width=.49\textwidth]{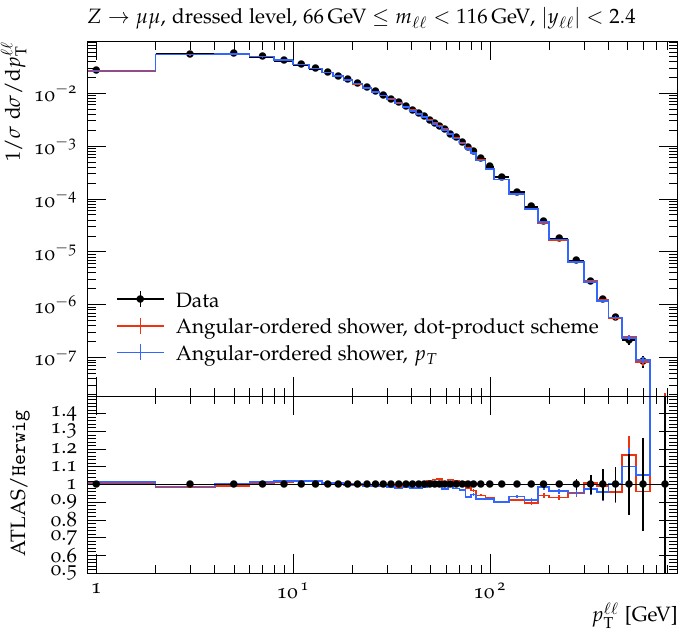}
\caption{Transverse momentum distribution of a $Z$ boson decaying into a pair of dressed muons in proton-proton collisions at $\sqrt{s}=8$~TeV.
  The data is from the ATLAS collaboration~\cite{ATLAS:2015iiu}.
  The red curve is given by the \Hw{} angular-ordered shower with the dot-preserving scheme (default as from version 7.3), while the
  blue curve uses the transverse-momentum ($p_T$) scheme.
  Both curves are obtained using Matrix-Element-Corrections to
  describe the hardest emission, and the shower parameters are those
  from Ref.~\cite{Bewick:2021nhc}.
}
\label{fig:pTZ_ATLAS8TeV_AO}
  \end{figure}

All three schemes are now available in \Hw~7.3 for both ISR and FSR,
with the dot-product preserving scheme the default.
The recoil scheme can be selected with the interface commands
\\[0.05in]
\texttt{cd /Herwig/Shower}\\
\texttt{set ShowerHandler:EvolutionScheme <scheme>}
\\[0.05in]
and if necessary
\\[0.05in]
\texttt{set PowhegShowerHandler:EvolutionScheme \hspace*{\fill}<scheme>}
\\[0.05in]
where \texttt{<scheme>} is \texttt{pT}, \texttt{Q2} or \texttt{DotProduct}.
We encourage the use of the transverse-momentum preserving scheme to
estimate shower uncertainties induced by the recoil scheme, but we
discourage the use of the virtuality-preserving scheme, as it
deteriorates the formal logarithmic accuracy of the angular-ordered
shower.
For this reason, the phenomenological study carried out in
Ref.~\cite{Bewick:2021nhc} does not include this latter scheme.
In Fig.~\ref{fig:pTZ_ATLAS8TeV_AO} we compare the transverse-momentum
distribution of the $Z$ boson produced in proton-proton collisions at
$\sqrt{s}=8$~TeV with the predictions obtained using the \Hw{}
angular-ordered shower for the dot-product~(red) and $p_T$~(blue)
recoil schemes.
We notice that the two schemes, which both reach next-to-leading
logarithmic accuracy for this observable, are in good agreement
between each other and with ATLAS data~\cite{ATLAS:2015iiu}.

\section{Developments of the cluster hadronization and secondary decay models}

\subsection{HQET in hadronization and decay of heavy and excited hadrons}

The enhancement of the transmission of heavy hadron polarization post-parton shower is pivotal for predicting heavy hadron behaviour in high-energy collisions. When the heavy quark mass $m_Q$ is significantly greater than the QCD scale $\Lambda_{\rm QCD}$, light degrees of freedom become insensitive to the mass, leading to crucial implications for the isospin heavy hadron multiplets ($H$, $H^\star$). In the context of HQET, it is found that the existence of a strong hierarchy between the ground state decay width $\Gamma(H \to X)$, multiplet  mass splitting $\Delta m$ and the decay width of the radiative decay $\gamma(H^\star \to H X)$:
\begin{align}
\Gamma(H \to X) \gg \Delta m \gg \gamma(H^\star \to H X),
\end{align}
could result in heavy quarks acting as static colour sources. This facilitates the emergence of a ``spin-flavour symmetry" for heavy quarks. Meanwhile, the Falk-Peskin ``no-win" theorem~\cite{Falk:1993rf} elucidates the limits of polarization information retrieval:
\begin{align*}
\Delta m \gg \Gamma \gg \gamma \quad \text{or} \quad  \Delta m \gg \gamma \gg \Gamma,
\end{align*}
implying that polarization details are lost in non-excited mesons under these conditions. This finding extends to other cases within similar hierarchies of mass splitting and decay widths.

The above-mentioned spin-flavour symmetry facilitates model-independent predictions for heavy hadron spectra and interactions, separating short-distance perturbative interactions from long-distance non-perturbative fragmentation processes. As a consequence, the attributes of the heavy quark~-- such as velocity, mass, and spin~-- remain largely unaffected by the dynamics of light quarks in the hadronization sequence. This separation of scales allows the preservation of heavy quark polarization, which could be reflected in the spin polarization of produced heavy mesons and baryons. However, this can be influenced by the light quark's angular momentum and parity-conserving fragmentation processes that may lead to anisotropies, characterized by a model-dependent parameter $\omega_j$.

For the $D_1$ and $D_2^{\star}$ meson states, considering the left-handed polarization of the charm quark, the light degrees of freedom with $j_q = \frac{3}{2}$ can exhibit any of the four helicity states. Due to parity invariance, the probability of forming a specific helicity state is independent of the helicity's sign but can vary for different helicity magnitudes. The parameter $\omega_j$ is introduced to denote the likelihood of fragmentation into a state with the maximum $|j_q^{(3)}|$ value. This allows for a breakdown of the probabilities for various helicity states, demonstrating how the combination of a left-handed charm quark with particular light quark helicities leads to distinctively populated helicity states with their respective probabilities for $D$, $D_0^{\star}$, $D_1$, and $D_2^{\star}$.
\begin{table*}[h]
\centering
\scriptsize
\begin{tabular}{|c||c c c c c|}
\hline
$\hat{\rho}$ & $\rho_{0,0}$ & $\rho_{1,1}$ & $\rho_{2,2}$ & $\rho_{3,3}$ & $\rho_{4,4}$ 
\\ \hline \hline
$D$ & $1$ & $-$ & $-$ & $-$ & $\begin{matrix}  \\ \\ \end{matrix} -\begin{matrix}  \\ \\ \end{matrix}$ 
\\
$D^{\star}$ & ${1\over 2}(1-\rho_Q)$
      & ${1\over 2}$
      & ${1\over 2}(1+\rho_Q)$
      & $-$
      & $\begin{matrix}  \\ \\ \end{matrix} -\begin{matrix}  \\ \\ \end{matrix}$ 
\\ 
$D_1$ & ${1\over 16}[1-\rho_Q + \omega_{3 \over 2}(3-5\rho_Q)]$
      & ${1 \over 4}(1-\omega_{3 \over 2})$
      & ${1\over 16}[1-\rho_Q + \omega_{3 \over 2}(3+5\rho_Q)]$
      & $-$
      & $\begin{matrix} \\ \\ \end{matrix}-\begin{matrix}  \\ \\ \end{matrix}$ 
\\ 
$D_2^{\star}$ & ${1 \over 4} \omega_{3 \over 2} (1-\rho_Q)$
      & $({3 \over 16} - {1 \over 8}\omega_{3 \over 2})(1-\rho_Q) $
      & ${1 \over 4} (1-\omega_{3 \over 2})$
      & $({3 \over 16} - {1 \over 8}\omega_{3 \over 2})(1+\rho_Q) $
      & $\begin{matrix} \\ \\ \end{matrix} 
        {1 \over 4} \omega_{3 \over 2} (1+\rho_Q)
        \begin{matrix} \\ \\ \end{matrix} $
\\ \hline
\end{tabular}
\caption{Polarization states of charmed mesons, $D$, $D^{\star}$, $D_1$ and $D_2^{\star}$~\cite{LE_PAPER}.}
\label{MesonPol}
\end{table*}

To determine the numerical value of $\omega_j$, we consider the amplitude for the production of a pion at $\theta,\phi$ from a $H^{\star} \to H\pi$ type meson decay, which is proportional to the spherical harmonics $Y_{j}^{\ell}(\theta, \phi)$ 
\begin{equation}
{d\Gamma(H^{\star} \to H\pi) \over d\cos\theta} \propto \int d \phi \sum_{j}  P_{H^{\star}}(j)  \bigl| Y_{j}^{\ell}(\theta,\phi) \bigr|^2,
\label{diff_width}
\end{equation} 
Here, $\ell$ is the angular momentum quantum number of $H^{\star}$ and $P_{H^{\star}}(j)$ is the probability of a $H^{\star}$ emerging with helicity $j$~\cite{LE_PAPER}. For instance, the differential decay width for $D_2^{\star} \to D\pi$ has been analyzed against experimental data, establishing an upper limit of $\omega_{3/2}<0.24$ at 90\% CL~\cite{Falk:1993rf}. In \Hw~7.3, we incorporate a \textcolor{blue}{\texttt{SpinHadronizer}} class, particularly its sub-function \texttt{mesonSpin}, to systematically assign spin and polarization to newly formed mesons based on the heavy quark's flavour and the meson's spin attributes, with a default value of $\omega_{3/2} = 0.20$~\cite{LE_PAPER}.

An analogous examination applies to heavy baryons. For these, the ground state is defined by a heavy quark paired with a light diquark system, which has a helicity of $j_{qq} = 0$. In such a state, no angular momentum is imparted to the heavy quark, thus the original polarization remains unaffected. Therefore, the initial polarization of the heavy quarks can be expected to directly affect the ground state polarization of heavy baryons. The likelihood of encountering these states during the fragmentation of heavy sectors continues to be dictated by two specific parameters, $\omega_a$ and $\omega_j$. Here, $\omega_a$ denotes the probability of forming a $j_{qq} = 1$ diquark in contrast to a $j_{qq} = 0$ ground state setup. Within the framework of \textsf{Herwig 7}'s cluster hadronization approach, $\omega_a$ is set to 1, reflecting an unbiased probability between the formation of spin-0 and spin-1 diquarks~\cite{LE_PAPER}. The numerical value of $\omega_1$ can be estimated similar to the case of $\omega_{3 \over 2}$, considering the decay widths of the observed decay modes $\Sigma_c \to \Lambda_c \pi$ and $\Sigma_c^{\star} \to \Lambda_c \pi$, which results in $\omega_{1} = {2/3}$. 

The above arguments allow one to determine the polarization distributions of the excited heavy mesons and heavy baryons, through the helicity state of their heavy quark constituents, as described in~\cite{LE_PAPER}. The explicit results for the case of excited heavy mesons and heavy baryons are tabulated in Tables~\ref{MesonPol} and \ref{BaryonPol}.

\begin{table*}
\centering
\begin{tabular}{|c | | c c c c|}
\hline
$\hat{\rho}$ & $\rho_{0,0}$ & $\rho_{1,1}$ & $\rho_{2,2}$ & $\rho_{3,3}$ 
\\ \hline \hline
$\Lambda_c$ & ${1\over 2}(1-\rho_Q)$
      & ${1\over 2}(1+\rho_Q)$
      & $-$
      & $\begin{matrix}  \\ \\ \end{matrix} -\begin{matrix}  \\ \\ \end{matrix}$ 
\\ 
$\Sigma_c$ & ${1\over 2}(1-\rho_Q) + \omega_{1} \rho_Q$
       & ${1\over 2}(1+\rho_Q) - \omega_{1} \rho_Q$
       & $-$
       & $\begin{matrix} \\ \\ \end{matrix}-\begin{matrix}  \\ \\ \end{matrix}$ 
\\ 
$\Sigma_c^\star$ & ${3\over 8}\omega_{1}(1-\rho_Q)$
      & ${1\over 2}(1-\rho_Q) - {1\over 8}\omega_{1}(3-5\rho_Q)$
      & ${1\over 2}(1-\rho_Q) - {1\over 8}\omega_{1}(3+5\rho_Q)$
      & $\begin{matrix} \\ \\ \end{matrix} 
        {3\over 8}\omega_{1}(1+\rho_Q)
        \begin{matrix} \\ \\ \end{matrix} $
\\ \hline
\end{tabular}
\caption{Possible polarization states of charmed baryons, $\Lambda_c$, $\Sigma_c$ and $\Sigma_c^\star$~\cite{LE_PAPER}.}
\label{BaryonPol}
\end{table*}

At this stage, it is crucial to note that due to the scarce experimental data around these heavy hadrons, we depend heavily on the principles of HQET to predict how they decay, including their lifetimes and decay modes, as it lays down fundamental connections between the coupling constants involved in these decays.  HQET has been already used for modeling the strong and radiative decay of heavy baryons, as shown in some prior studies and its integration into \HWPP~\cite{Bahr:2008pv}. Nevertheless, we still have to develop such a prescription for the heavy meson sector. We begin by identifying the $s-$ and $p-$wave meson multiplets: the ground state doublets ($J^P=0^-, 1^-$) including $D$ and $D^{\star}$ mesons, the $J^P=1^+, 2^+$ doublet with $D_1$ and $D_2^{\star}$ mesons, and the $J^P=0^+, 1^+$ doublet comprising $D_0^{\star}$ and $D_1^\prime$ mesons \cite{Falk:1992cx}. We also account for the mixing effects between $D_1$ and $D_1^\prime$ mesons resulting from higher-order corrections in HQET. The methodology and notation for calculating the matrix elements associated with meson decays are as per Falk et al. \cite{Falk:1992cx,Falk:1995th}, with a focus on balancing theoretical precision with empirical data representation. The matrix elements are defined by equations (\ref{m1}) through (\ref{m6}) below, incorporating decay parameters and polarization vectors, where $p_i$ denotes the hadron momenta and $\epsilon_i$ the polarization vectors \cite{LE_PAPER}.
Here, the index $i=0$ identifies the parent hadron, while $i=1$ and $i=2$ refer to the resulting heavy and light daughter hadrons. The mass of a hadron $H$ is represented by $m_H$. The parameters $g$, $h$, $\Lambda$, $f_{\pi}$ and $f''$ are constants associated with the decay process. Furthermore, $\theta_q$ is introduced as the mixing angle between the meson pairs ($D_1$, $D'_1$) and ($D_{s1}$, $D'_{s1}$).
\begin{strip}
\begin{align}
\mathcal{M}(D^{\star}\to D\pi) &= -{2g \over f_\pi} \left( m_Dm_{D^{\star}} \right)^{1\over 2} p_0\cdot \epsilon_0 ,
\label{m1}
\\
\mathcal{M}(D_2^{\star}\to D\pi) &= -{2h \over f_\pi \Lambda} \left( m_{D_2}m_D^{\star} \right)^{1\over 2} \epsilon_0^{\mu\nu} p_{0,\mu} p_{0,\nu} ,
\label{m2}
\\
\mathcal{M}(D_2^{\star}\to D^{\star}\pi) &= -i{2h \over f_\pi \Lambda} \left( {m_{D^{\star}} \over m_{D_2}} \right)^{1\over 2}  
\epsilon^{\alpha\beta\mu\nu} \epsilon^0_{\alpha\gamma} p_0^\gamma p_{0,\mu} p_{1\nu}\epsilon_{1\beta} ,
\label{m3}
\\
\mathcal{M}(D_1\to D^{\star}\pi) &=  {h \over f_\pi\Lambda}   \left( {2 \over 3} {m_{D_1}m_D} \right)^{1\over 2}
\bigg[ \epsilon_0\cdot\epsilon_1\left(p_0^2-\left[{p_0\cdot p_1 \over m_0}\right]^2\right)
-3\epsilon_0\cdot p_0\epsilon_1\cdot p_0 \bigg] ,
\label{m4}
\\
\mathcal{M}(D_0^{\star}\to D\pi) &= {f^{\prime\prime} \over f_\pi}\left( {m_{D_0^{\star}}m_D} \right)^{1\over 2} \,p_0\cdot
\left({p_1 \over m_{D^{\star}_0}}+{p_2 \over m_D}\right) ,
\label{m5}
\\
\mathcal{M}(D_1^\prime \to D^{\star}\pi) &=-{f^{\prime\prime} \over f_\pi}\left( {m_{D_1^\prime}m_D} \right)^{1\over 2}
\bigg[-p_0\cdot\left({p_1 \over m_{D^{\star}_0}}+{p_2 \over m_D}\right)\epsilon_0\cdot\epsilon_1 
+{1 \over m_{D'_1}}\epsilon_1\cdot p_1\epsilon_0\cdot p_0
+{1 \over m_{D}}\epsilon_0\cdot p_2\epsilon_1\cdot p_0 \bigg].
\label{m6}
\end{align}
\end{strip}

With the established matrix elements for the decays at hand, we are equipped to determine the partial decay widths of the relevant two-body decay processes. To align our theoretical framework with the practical applications within \textsf{Herwig 7}, it is imperative to consider certain non-dominant terms that become significant in the context of the heavy quark expansion. In practical terms, this is described by:
\begin{align}
\Gamma(H^{*} \rightarrow H \pi) = \frac{1}{8\pi m_{H^{*}}^2} 
\left| \mathcal{M}(H^{*} \rightarrow H \pi) \right|^2 p_{\text{CM}},
\end{align}
where $p_{\rm CM}$ is the momentum of the decaying particle in the center-of-mass system of the decayed two-body system. The resulting partial widths are reported in~\cite{LE_PAPER}.

We calculated the decay parameters using recent measurements of charmed meson masses and decay widths. These parameters are important for understanding how these particles transform into others over time. Table~\ref{tab7} lists the best-fit values for these parameters.
\begin{table}[h]
\centering
\begin{tabular}{|c||c|}
\hline 
Parameter      & Fitted Value              \\ \hline \hline
$f''$          & $ -0.465 \pm 0.017$       \\ \hline
$f_\pi$        & $  0.130 \pm 0.001$ [GeV] \\ \hline
$h$            & $  0.824 \pm 0.007$       \\ \hline
$\Lambda$      & $  1.000 \pm 0.000$ [GeV] \\ \hline
$g$            & $  0.565 \pm 0.006$       \\ \hline
$\theta_{u,d}$ & $  0.000 \pm 0.100$       \\ \hline
$\theta_s$     & $ -0.047 \pm 0.002$       \\ \hline
\end{tabular}
\caption{Fitted values of the decay parameters.}
\label{tab7}
\end{table}
Note that some mesons have decay modes that do not conserve isospin strongly. These modes become important when the usual isospin-conserving decays are not allowed or are very unlikely. We pay particular attention to the $D^{\star}$ mesons, which have various decay possibilities because of their energy levels. Our methods for examining these decays can also be applied to other mesons~\cite{LE_PAPER}.

\begin{figure}[t]
\centering
\includegraphics[width=.49\textwidth]{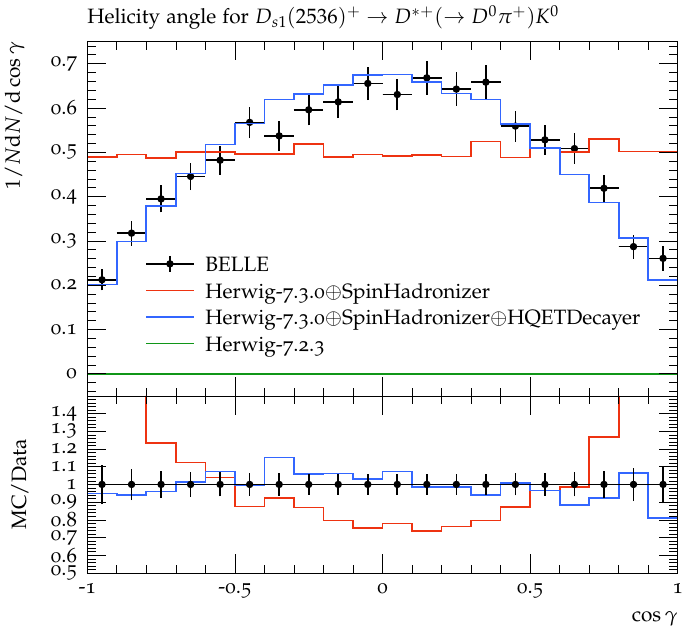}
\caption{Efficiency-corrected decay rates for $D_{s1} \to D^{\star +} K^0$ decay mode as a function of the angle ($\gamma$) between $\pi^+$ and $K^0$ in the $D^{\star +}$ rest frame. The data is from BELLE collaboration~\cite{Belle:2007kff}. The plot is from~\cite{LE_PAPER}.}
\label{SpinHadrBELLE}
\end{figure}

To integrate the processes of strong and radiative decays for excited heavy mesons into \Hw~7, we have created two dedicated classes: \textcolor{blue}{\texttt{HQETStrongDecayer}} and \textcolor{blue}{\texttt{HQETRadiativeDecayer}}. The decay parameters, emphasized in this segment, are set as variables that users can modify. Such an arrangement offers adaptability, allowing for adjustments and enhancements in line with new findings or needs. The efficacy of using HQET and spin-flavour symmetry to predict polarization-sensitive measurements has been extensively discussed in~\cite{LE_PAPER}, see e.g. Figure~\ref{SpinHadrBELLE}. 

\subsection{Improved cluster splitting dynamics}
\label{ref:cluster}
In \Hw~7's cluster hadronization model, the dynamics of hadron formation are conceptualized as a multi-step process, beginning with fragmentation and culminating in the production of observable hadrons. Initially, during fragmentation, coloured partons, produced in the primary hard process and the subsequent parton shower, are paired to form colour-neutral clusters. This pairing typically involves a quark with an antiquark, often originating from the breaking of colour flux tubes via the insertion of quark-antiquark pairs. Once formed, these clusters represent temporary and unstable combinations of partons. The subsequent step involves the decay or splitting of these clusters into final-state hadrons. For lighter clusters, this might involve a direct transformation into a meson or baryon. However, heavier clusters typically undergo a series of successive splittings until they reach configurations suitable for direct conversion into observable hadrons. This is handled by the \textcolor{blue}{\texttt{ClusterFissioner}} class.  

Each of these cluster splittings or decays is governed by the kinematic checks over the parent and children clusters, primarily focusing on the relation between the masses of the clusters and their constituent quarks. Formerly, \Hw~7.2 and older versions used only a static kinematic threshold for 2-body cluster splittings
\begin{align}
M_0 > M_1 + M_2, \quad M_1 > m + m_1, \quad M_2 > m + m_2,
\label{static}
\end{align}
with $M_i$ being the masses of the parent and children clusters. $m_{1,2}$ are the masses of the parent cluster's constituent quarks and $m$ is the quark mass of a perturbatively spawned light quark-antiquark pair. Aiming to improve the kinematics of cluster splitting, \Hw~7.3 by default employs a new dynamic kinematic threshold scheme, in addition to keeping the static option available:
\\[0.05in]
\texttt{cd /Herwig/Hadronization}\\
\texttt{set  ClusterFissioner:KinematicThreshold \hspace*{\fill}<Static/Dynamic>}
\\[0.05in]
The \texttt{Dynamic} choice updates the condition~\eqref{static} with~\cite{LE_PAPER}
\begin{align}
M_0^2 > M_1^2 &+ M_2^2 , \quad M_1^2 > m^2 + m_1^2 + \delta_{\rm th}, \nonumber \\ &M_2^2 > m^2 + m_2^2 + \delta_{\rm th},
\label{dynamic}
\end{align}
where the \texttt{KineticThresholdShift} parameter $\delta_{\rm th}$ allows one to adjust this threshold through tuning. In this new regime,  satisfaction of the condition~\eqref{dynamic} is not sufficient to allow for a cluster to split; an additional decision-making device in the form of a scale-dependent threshold-modulated probability distribution has been introduced:
\begin{align}
P_{\rm cluster} = \frac{1}{1 + \left| (M - \delta)/M_{\rm th} \right|^r} > {\rm Random}[0,1],
\label{bell}
\end{align}
whereas $M_{\rm th}$ signifies the cluster's mass threshold~-- the aggregate of the masses of the inherent quarks and the produced di-quark. The splitting is allowed only if $P_{\rm cluster}$ is greater than a random number between $[0,1]$. The parameters $\delta$ and $r$ may be chosen in a way to reduce the chance of splitting in successive instances. 

Collectively, the addition of these three tunable parameters, namely \texttt{ProbabilityPowerFactor} ($r$), \texttt{ProbabilityShift} ($\delta$), and \texttt{KineticThresholdShift} ($\delta_{\rm th}$) allows the user to exact further control over the cluster splittings~\cite{LE_PAPER}. This significant change required a new tune of \Hw~7.3 to the data, which will be addressed in Section~\ref{Gtune}.

\subsection{New infrastructure for hadronization models}

Several improvements are in development with respect to the cluster
hadronization and will soon be available with an upcoming Herwig
release, among them a dynamic cluster model
\cite{Hoang:2024zwl} and other options for an improved
understanding of hadronization. Many of the structures already appear
in this release and allow for the flexible adjustment of kinematics
and mass distributions in the cluster fission process (offering more
alternatives to the choices made in Sec.~\ref{ref:cluster}), as well
as the generation of dynamic gluon masses using the
\textcolor{blue}{\texttt{GluonMassGenerator}} class.

\subsection{Event-by-event hadronization corrections}

As part of the hadronization improvements, a new strategy has been
developed to transfer the assignment of constituent masses entirely
into the hadronization. Not only does this allow to have a consistent
physics interface to the string model, but it also provides the
possibility to extract event-by-event hadronization corrections in a
clean way, {\it i.e.} a parton level which is not 'contaminated' by
the non-perturbative constituent mass parameters. This is achieved
by reshuffling the partonic ensemble to different mass shells at the
beginning of the hadronization. While this can, in principle, be done
across the entire event, the more physical choice is to perform the
reshuffling within colour singlet subsystems which will branch into
clusters:
\vspace*{1ex}

\noindent\texttt{cd /Herwig/Hadronization}\\
\texttt{set ClusterHadHandler:Reshuffle Yes} \\
\texttt{set ClusterHadHandler:ReshuffleMode \hspace*{\fill}ColourConnected} \\
\texttt{cd /Herwig/Shower}\\
\texttt{set ShowerHandler:UseConstituentMasses No}
\vspace*{1ex}

The intermediate partonic state is then tagged by a status code which
can be chosen at the level of the input file, and is available for
analysis by reading it out from the event record:
\vspace*{1ex}

\noindent\texttt{cd /Herwig/Shower}\\
\texttt{set ShowerHandler:TagIntermediates \\\hspace*{\fill}<status code>}

\subsection{Particle data update}

In \Hw-7.3.0, we have undertaken the first substantial update of particle data since the initial launch of \Hw~7. This update incorporates the latest findings from the Particle Data Group (PDG) 2022 review~\cite{ParticleDataGroup:2022pth}, marking a significant enhancement from the previously used PDG 2006 data~\cite{ParticleDataGroup:2006fqo}. Updated elements include particle masses, decay widths, decay modes, and branching ratios, ensuring that the users have access to simulations that employ the most up-to-date and accurate particle physics data available.

\subsection{String Hadronization tune with colour reconnection}
With the new version of \Hw, we add the possibility of using the Lund string model 
for hadronization in both $e^+e^-$ and $pp$ collisions. The interface of the string model of 
\Pythia{} is provided via the \PythiaI{} C++ package (written by L.~Lönnblad) \cite{PIeight}. 
While the default version of the \PythiaI{} is sufficient for hadronization
of electron-positron collisions, we had to extend it to account 
for Colour Reconnection which is needed for realistic
simulation of hadron-hadron collisions.
The Lund string model, together with the angular ordered parton shower model,
has been tuned to LEP and LHC data sets. Therefore,
it is now possible to use the angular ordered parton shower of \Hw{} together with two different hadronization models, which may help estimate the differences associated with hadronization effects. Details will be described in a forthcoming publication.

\section{\Hw~7.3 General Tune}
\label{Gtune}

As with previous versions, we have tuned \Hw~7.3 using $e^+e^-$ data from the LEP, PETRA, SLAC, SLC and TRISTAN measurements for over 9,200 individual data bins, weighted around both light and heavy hadron production rates and multiplicities, alongside a number of dominant processes. We have explored 12 parameters in total, with 10 related to cluster hadronization and the remaining 2, specifically \texttt{AlphaIn} and \texttt{pTmin}, associated with the AO parton shower. Due to the large number of relevant parameters and the sensitivity of their collective phase space, we opted for a multi-layered, brute-force tuning approach employing the \texttt{prof2-chisq} module of \textsf{ProfessorII}~\cite{Buckley:2009bj}, aiming to minimize $\chi^2$ value as a benchmark for optimal tuning~\cite{LE_PAPER}. 

The first layer (1000 samples) was probed around the whole viable phase space, with the lower threshold chosen to be the $\chi^2$ of the tuned \Hw-7.2.3. The benchmark closest to the threshold was then chosen for the second layer, with the relative sampling phase space being tightened around it by 50\%. After five layers of successive sampling, our strategy delivered a marked enhancement in the overall $\chi^2$ value. Specifically, the tuned \Hw~7.3 achieved a roughly 50\% $\chi^2$ improvement over \Hw~7.2 and approximately 13\% improvement against \textsf{Herwig-7.2.3}. The parameter content and the result of this new general tune for \Hw~7.3 is tabulated in Table~\ref{tuned_parameters}.
\begin{table*}[h]
\centering
\scriptsize
\begin{tabular}{|c||l||c c|}
 \hline
Tuned Parameter & Description & \textsf{Herwig-7.3.0} & \textsf{Herwig-7.2.0} \\
\hline\hline
\texttt{ClMaxLight} [GeV]& Maximum allowable cluster mass for light quarks. & 3.529 & 3.649 \\
\texttt{ClPowLight} & Power exponent for the mass of clusters with light quarks. & 1.849 & 2.780 \\
\texttt{PSplitLight} & Parameter affecting the mass splitting for clusters with light quarks. & 0.914 & 0.899 \\
\texttt{PwtSquark} & Probability for a $s\bar{s}$ quark pair to be spawned during cluster splittings. & 0.374 & 0.292 \\
\texttt{PwtDIquark} & Probability for quarks forming a di-quark. & 0.331 & 0.298 \\
\texttt{SngWt} & Weighting factor for singlet baryons in hadronization. & 0.891 & 0.740 \\
\texttt{DecWt} & Weighting factor for decuplet baryons in hadronization. & 0.416 & 0.620 \\
\texttt{ProbabilityPowerFactor} & Exponential factor in the \texttt{ClusterFissioner} probability function. & 6.486 & $-$ \\
\texttt{ProbabilityShift} & Offset in the \texttt{ClusterFissioner} probability function. & -0.879 & $-$ \\
\texttt{KineticThresholdShift} [GeV]& Adjustment to the kinetic threshold in \texttt{ClusterFissioner}. & 0.088 & $-$ \\
\texttt{AlphaIn} & Initial value for the strong coupling constant at $M_{Z^0}=91.1876$ GeV. & 0.102 & 0.126 \\
\texttt{pTmin} [GeV]& Minimum transverse momentum in parton shower. & 0.655 & 0.958 \\
\hline
\end{tabular}
\caption{Tuned parameters in \Hw~7.3 vs \Hw~7.2, as outlined in~\cite{LE_PAPER}.}
\label{tuned_parameters}
\end{table*}

A retuning of parameters affecting underlying event and minimum bias
collisions is currently ongoing.

\section{Other Changes}

Besides the major physics improvements highlighted in the previous
sections, we have also made a number of smaller changes to the code
and build system. Please refer to the
online documentation for a fully detailed description or contact the
authors.

Recently, the first steps have been taken towards creating \HADML{} a machine learning hadronization model~\cite{Ghosh:2022zdz} and a protocol for fitting it to unbinned experimental data~\cite{Chan:2023ume}.  Although the model is still incomplete it has been successfully interfaced with \Hw. Once the model is finalized, it is planned to release it with \Hw.

\section{Example Results}

\Hw~7.3 has been thoroughly validated against a wide range of existing
data, as implemented in the Rivet and FastJet
frameworks\cite{Buckley:2010ar,Cacciari:2011ma}. Parameter tuning has
been performed using Professor\cite{Buckley:2009bj}.

A wide range of further plots can be found at
\texttt{\href{https://herwig.hepforge.org/plots/herwig7.3}{https://herwig.hepforge.org/plots/herwig7.3}}.

\section{Summary and Outlook}

We have described a new release, version 7.3, of the \Hw\ event
generator. This new release contains a number of improvements to both
perturbative and non-perturbative simulation of collider physics and will form
the basis of further improvements to both physics and technical aspects.

\section*{Acknowledgments}

We are indebted to Leif L\"onnblad for his authorship of \ThePEG, on which
\Hw\ is built, and his close collaboration, and to the authors of
\Rivet,  \PythiaI{} and \textsf{Professor}. We are also grateful to Malin Sj\"odahl for
providing the \textsf{ColorFull} library for distribution along with \Hw.

This work was supported in part by the European Union as part of the 
H2020 Marie Sk\l odowska-Curie Initial Training Networks MCnetITN3
(grant agreement no.\ 722104) and the UK Science and Technology Facilities Council (grant numbers
ST/T001011/1, ST/T001038/1). 
A.P.\ acknowledges support by the National Science Foundation under Grant No.\ PHY 2210161.
G.B.\ thanks the UK Science and Technology Facilities Council for the award of a studentship.
The work of A.S.\ and J.W.\ was funded by grant no.\ 2019/34/E/ST2/00457 of the National Science
Centre, Poland. A.S.\ was also supported by the Priority Research Area Digiworld under the program Excellence Initiative – Research University at the Jagiellonian University in Cracow.
This work has been supported by the BMBF under grant number 05H21VKCCA.

\bibliography{Herwig}
\end{document}